\documentclass[abbrv,aps,prl,twocolumn,showpacs,preprintnumbers,amsmath,amssymb,
superscriptaddress]{revtex4}

\usepackage{graphicx}
\usepackage{dcolumn}
\usepackage{bm}
\usepackage{color}

\makeatletter
\newcommand*{\balancecolsandclearpage}{
  \close@column@grid
  \clearpage
  \twocolumngrid
}
\makeatother

\begin{document}
\title {Observation of Dirac plasmons in a topological insulator}

\author{P. Di Pietro} 
\affiliation {CNR-SPIN and Dipartimento di Fisica, Universit\`a di Roma "La Sapienza",
Piazzale A. Moro 2, I-00185 Roma, Italy}

\author{M. Ortolani} 
\affiliation {CNR-IFN and Dipartimento di Fisica, Universit\`a di Roma "La Sapienza",
Piazzale A. Moro 2, I-00185 Roma, Italy}

\author{O. Limaj} 
\affiliation {Dipartimento di Fisica, Universit\`a di Roma "La Sapienza" and INFN,
Piazzale A. Moro 2, I-00185 Roma, Italy}

\author{A. Di Gaspare} 
\affiliation {CNR-IFN, Via Cineto Romano, 00000 Roma}

\author{V. Giliberti} 
\affiliation {CNR-IFN and Dipartimento di Fisica, Universit\`a di Roma "La Sapienza",
Piazzale A. Moro 2, I-00185 Roma, Italy}

\author{F. Giorgianni} 
\affiliation {Dipartimento di Fisica, Universit\`a di Roma "La Sapienza" and INFN,
Piazzale A. Moro 2, I-00185 Roma, Italy}
  
\author{M. Brahlek} 
\affiliation{ Department of Physics and Astronomy Rutgers, The State University of New Jersey 136 Frelinghuysen Road Piscataway, NJ 08854-8019 USA}

\author{N. Bansal}
\affiliation{ Department of Physics and Astronomy Rutgers, The State University of New Jersey 136 Frelinghuysen Road Piscataway, NJ 08854-8019 USA}

\author{N. Koirala}
\affiliation{ Department of Physics and Astronomy Rutgers, The State University of New Jersey 136 Frelinghuysen Road Piscataway, NJ 08854-8019 USA}

\author{S. Oh}
\affiliation{ Department of Physics and Astronomy Rutgers, The State University of New Jersey 136 Frelinghuysen Road Piscataway, NJ 08854-8019 USA} 

\author{P. Calvani} 
\affiliation {CNR-SPIN and Dipartimento di Fisica, Universit\`a di Roma "La Sapienza",
Piazzale A. Moro 2, I-00185 Roma, Italy}

\author{S. Lupi}
\affiliation {CNR-IOM and Dipartimento di Fisica, Universit\`a di Roma "La Sapienza",
Piazzale A. Moro 2, I-00185 Roma, Italy}

\pacs{71.30.+h, 78.30.-j, 62.50.+p}

\maketitle

\textbf{Plasmons are the quantized collective oscillations of electrons in metals and doped semiconductors. The plasmons of ordinary, massive electrons are since a long time basic ingredients of research in plasmonics and in optical metamaterials \cite{Maier}.
Plasmons of massless Dirac electrons were instead recently observed in a purely two-dimensional electron system (2DEG) like graphene \cite{Martin}, and their properties are promising for new tunable plasmonic metamaterials in the terahertz and the mid-infrared frequency range \cite{Grigorenko}.
Dirac quasi-particles are known to exist also in the two-dimensional electron gas which forms at the surface of topological insulators due to a strong spin-orbit interaction \cite{Hasan1}. Therefore, one may look for their collective excitations by using infrared spectroscopy. Here we first report evidence of plasmonic excitations in a topological insulator (Bi$_2$Se$_3$), that was  engineered in thin micro-ribbon arrays of different width $W$ and period $2W$  to select suitable values of the plasmon wavevector $k$. Their lineshape was found to be extremely robust vs. temperature between 6 and 300 K, as one may expect for the excitations of topological carriers. Moreover, by changing $W$ and measuring in the terahertz range the plasmonic frequency $\nu_{P}$ vs. $k$  we could show, without using any fitting parameter, that the dispersion curve is in quantitative agreement with that predicted for Dirac plasmons.}

A topological insulator (TI) is a  quantum electronic material with an insulating gap in the bulk, of spin-orbit origin, and gapless  surface states at the interface with the vacuum or another dielectric. The latter states are metallic and associated with massless Dirac quasi-particles \cite{Hasan1, Kane1, Moore1}. The transport properties of these states, which exhibit chirality, are protected from back-scattering by the time-reversal symmetry and cannot be destroyed or gapped by  scattering processes which do not involve magnetic impurities.
Since their discovery TI's raised great interest, not only for their outstanding physical properties, like \textit{axionic} electromagnetic response \cite{Qi, Essin}, and exotic superconductivity \cite{Fu, Nilsonn}, but also for the potential applications in quantum computing \cite{Fu1,Kitaev1}, terahertz (THz) detectors \cite{zangh} and spintronic devices \cite{Chen}.  Like for other compounds, some of these foreseen applications may benefit from the exploitation of the 2DEG collective excitations, namely from plasmonics. Among the 2DEGs, the TI surface states  present the advantage that they spontaneously provide a two-dimensional Dirac system from the bulk material, without physically implementing an atomic monolayer like in graphene.  Moreover, thanks to the momentum-spin lockage, TI plasmons may potentially preserve the coherence of the electronic states up to room temperature. This would be a major step forward in quantum mechanics applications.

\begin{figure}[h]   
\begin{center}    
\includegraphics[width=18cm]{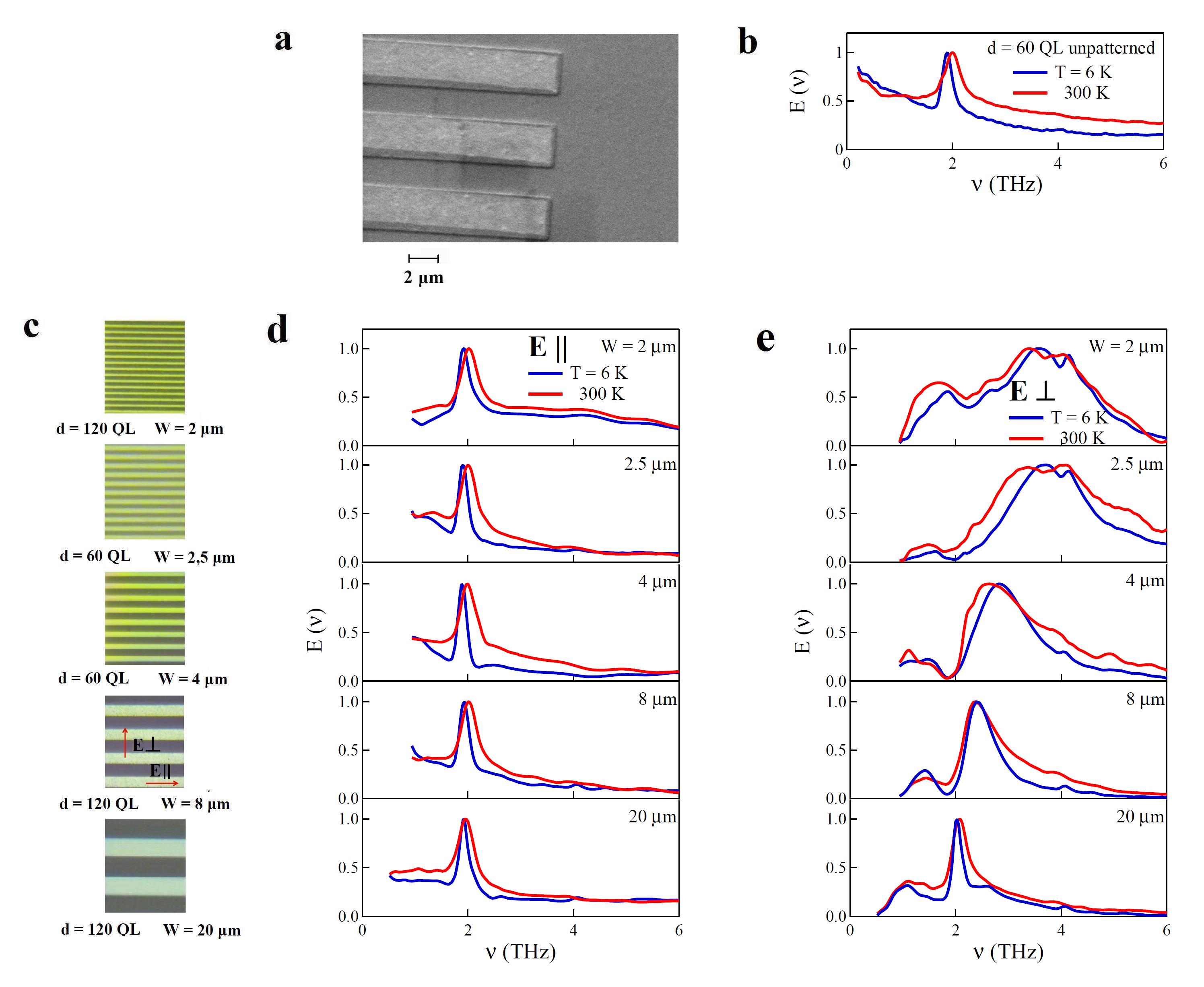}
\caption{}
\label{ribbons}
\end{center}
\end{figure}

In the 2DEG at the TI surface, collective excitations (plasmons) like those recently detected in graphene \cite{Martin} are indeed expected to exist. However, they cannot be directly excited by electromagnetic radiation because their dispersion law is such as to prevent the conservation of momentum in the photon absorption process. In other 2D systems, the necessary extra-momentum was provided through a patterning of the surface with a subwavelength grating \cite{Martin, Allen}. Here we have applied this methodology to thin films of Bi$_2$Se$_3$. They were patterned in form of micro-ribbon arrays of different widths $W$ and periods $2W$ (Fig. \ref{ribbons}-a,c). 
We could thus detect, in  the THz range, the optical absorption from collective oscillations of the electrons confined within the TI ribbons, namely,  the plasmons of the topological insulator. Their 2D character was confirmed by varying $W$  and measuring the corresponding peak frequency $\nu_{P}$, as the plasmon wavevector is related to $W$ by  $k \simeq \pi/W$ \cite{Nikitin1,Nikitin2}. We  thus found the expected two-dimensional dispersion law $\nu_{P} \propto \sqrt{k}$. Finally, we could assign the plasmon to massless Dirac electrons by observing its robustness against temperature changes from 6 to 300 K and by a comparison with theoretical calculations. Indeed, the experimental dispersion law was exactly reproduced by that of a 2D Dirac plasmon \cite{Polini, DasSarma} without using any fitting parameter, but just using the experimental values for the Fermi velocity $v_F$, the 2D charge density $n_{D}$, and the dielectric constant $\epsilon$, measured in Bi$_2$Se$_3$ films grown in the same experimental conditions \cite{Bansal, Bansal1}.

Six thin films of Bi$_2$Se$_3$ were grown by  Molecular Beam Epitaxy on 0.5 mm thick sapphire substrates  (Al$_2$O$_3$) \cite{Bansal, Bansal1}. Three out of them had a thickness $d$ = 120 quintuple layers (QL),  where 1 QL $\simeq$ 1 nm, the other three $d$ = 60 QL. Transport characterization through resistivity and Hall measurements show that  both Dirac electrons generated by topology and bulk massive electrons due to band-bending effect \cite{Bansal, Bansal1} participate to the surface conduction. 
One film was kept as grown for sake of comparison, the other five were  patterned by electron-beam lithography and reactive-ion etching in form of parallel ribbons of widths $W$  =2, 2.5, 4, 8, 20 $\mu$m, and periods $L=2W$. Therefore, the filling factor was 0.5  for all patterned samples (Fig.1-a, c). 

The transmittance $T(\nu)$ of the six films was measured in the THz range by a Fourier-transform interferometer from 6 to 300 K. The  corresponding extinction coefficients $E(\nu) = 1-T(\nu)$ are reported in Fig \ref{ribbons}-b for the as-grown sample and for the patterned films (d, e), once normalized by the respective peak values, both at 6 K (blue lines) and 300 K (red lines). 
The as-grown film in Fig. \ref{ribbons}-b exhibits  two peaks at the frequencies already reported for Bi$_2$Se$_3$ single crystals, namely, the $\alpha$ phonon mode  at 1.85 THz (61 cm$^{-1}$)  and the barely discernible $\beta$ phonon mode at  4.0 THz (132 cm$^{-1}$) \cite{Dipietro} which broaden at 300 K. Phonon lines are superimposed to a Drude absorption, which was mainly attributed to Dirac surface states \cite{Armitage}. 

The extinction coefficients of the patterned samples are reported in Fig \ref{ribbons}-d for the radiation field parallel to the ribbons. In this case the absorption, and its $T$-dependence, are very similar to those of the non-patterned film. This comparison also shows that  the patterning procedure did not affect at all the physical properties of the samples (see also Table S1 of the Supplementary Information). Moreover, in  Fig. \ref{ribbons}-d one may remark  that the phonon frequencies  neither appreciably change with $W$, nor with $d$.  

Figure \ref{ribbons}-e shows instead the extinction coefficient for the radiation polarized perpendicularly to the ribbons. As this direction is that of the reciprocal-lattice vectors needed for the energy-momentum conservation in the plasmonic absorption, it is  the one where  the plasmon can be observed. As one can see, either at 6 K and at 300 K  the $\alpha$ phonon is replaced by  a double absorption, where both peak frequencies strongly depend on $W$. We assign these features to the $\alpha$ phonon and to the plasmon of Bi$_2$Se$_3$, mutually interacting via a Fano interference. This produces a renormalization of both the phonon and the plasmon frequency, with a hardening of the mode at higher frequency and a softening of that at lower frequency, independently of their nature. A similar effect is reported in the literature for doped graphene on SiO$_2$ \cite{Basov}.  A Fano effect on the weak $\beta$ phonon line is also observed in the top panel of  Fig. \ref{ribbons}-e, as it becomes closer to the plasmon resonance, through an inflection point in $E(\nu)$ at the bare phonon frequency (magenta line).

In order to  extract from the data in Fig. \ref{ribbons}-e the bare plasmon ($\nu_{P}$) and phonon ($\nu_{ph}$) frequency we have fit the experimental data to the following Equation, obtained by Giannini et al. \cite{giannini}: 

\begin{equation}
E(\nu')=\frac{(\mathcal{\nu'}+q(\nu'))^{2}}{\mathcal{\nu'}^{2}+1}\cdot \frac{g^2}{1+\left(\frac{\nu-\nu_{P}}{\Gamma_{P}/2}\right)^2}
\label{giannini_1}
\end{equation}

\noindent The expressions for the renormalized frequency $\nu'$ which depends on $\nu_{ph}$ and for the (Fano)  parameter $q$,  that is the ratio between the  probability amplitude of exciting a discrete state (phonon) and of exciting a continuum or quasi-continuum state (plasmon),  are reported in the Methods section. $\Gamma_P$ is the plasmon linewidth and $g$ is the coupling factor of the radiation with the plasmon. Equation \ref{giannini_1} takes into account the frequency separation between the plasmon and the phonon excitations which here changes with $W$. 

\begin{figure}[h]   
\begin{center}    
\includegraphics[width=12cm]{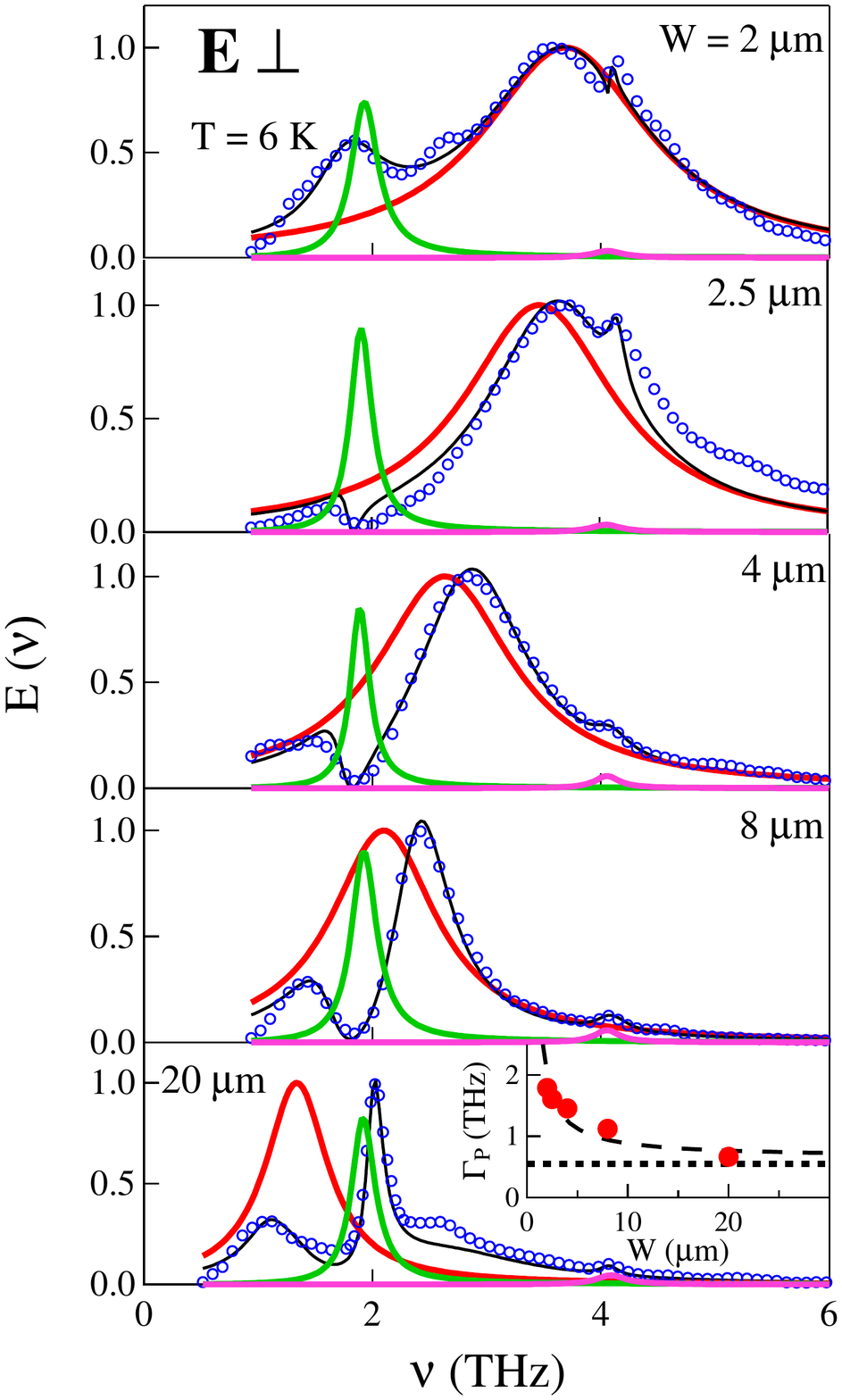}
\caption{}
\label{bare_plasmon}
\end{center}
\end{figure}

The above model satisfactorily reproduces the experimental data as shown in Fig. \ref{bare_plasmon} by the black line. The green and red lines describe instead the bare phonon and plasmon contributions, respectively, reconstructed in terms of a Lorentzian shape. As one can see, the bare phonon line does not change with $W$ and its frequency is the same as for the parallel polarization in Fig. \ref{ribbons}-d, while the plasmon softens and narrows as $W$ increases. This effect is better shown in the inset of Fig. \ref{bare_plasmon} (bottom panel) where the plasmon linewidth $\Gamma_P $ is shown vs. $W$ at 6 K. $\Gamma_P$ can be assumed to be the sum of the following independent contributions: i) the Drude linewidth obtained by fitting the spectra of Fig. \ref{ribbons}-d (see Table S1 in SI);  ii) the Landau damping rate, through both the creation of hole-electron pairs, and a phonon-assisted process \cite{Yan}; iii) radiative decay into photons; iv) finite size effects \cite{Yan}. Contribution i) is independent of $W$ and provides the background marked by the black dotted line in the inset. The other effects are expected to increase with increasing plasmon frequency (decreasing $W$) and therefore should be responsible for the behavior of $\Gamma_P$ shown in the inset.

\begin{figure}[h]   
\begin{center}    
\includegraphics[width=12cm]{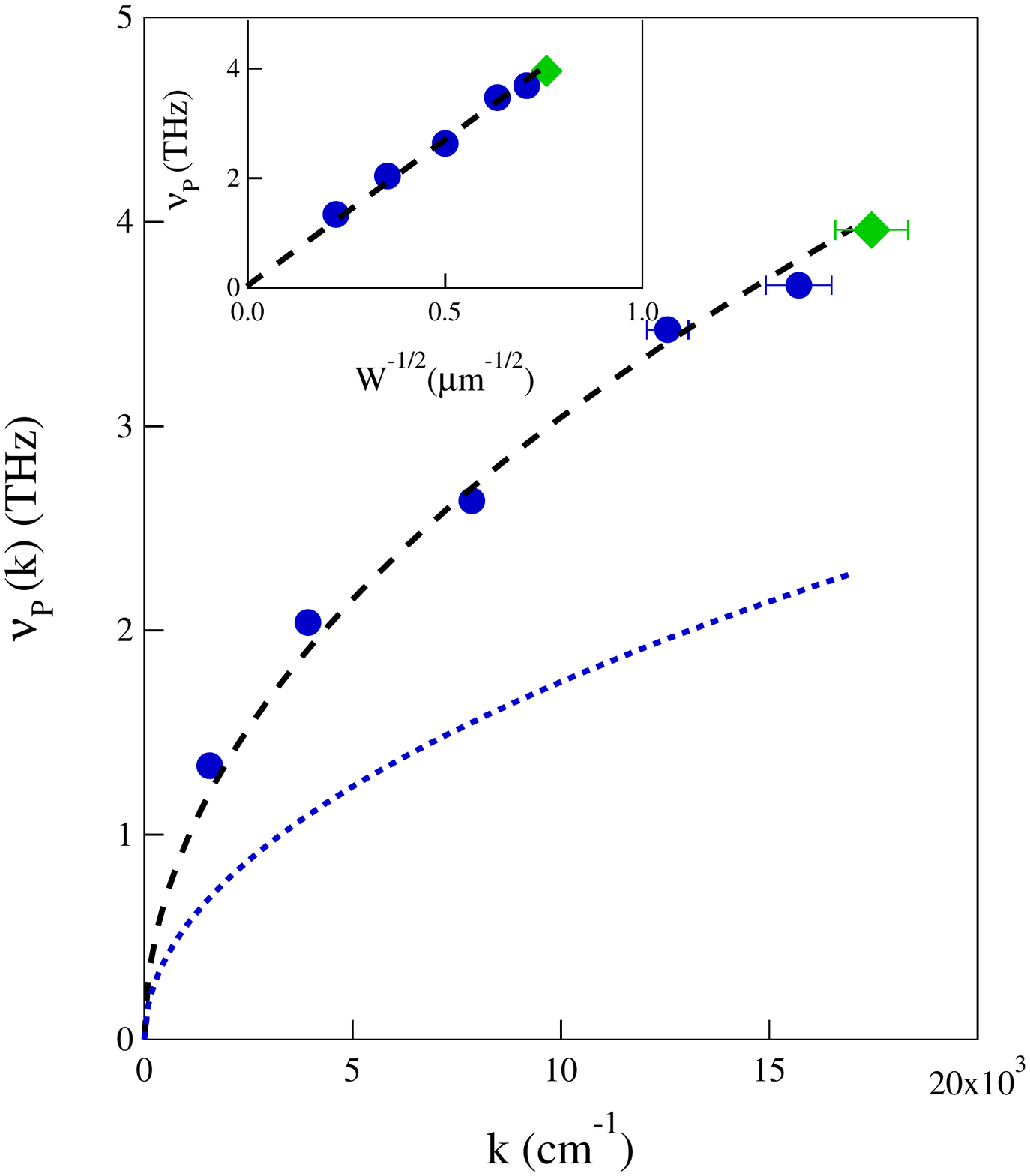}
\caption{}
\label{dispersion}
\end{center}
\end{figure}

The bare plasmon frequencies obtained from the Fano fits are plotted vs. the wavevector $k$ in Fig. \ref{dispersion} and vs. $W^{-1/2}$ in its inset. The additional point (green diamond) refers to a seventh sample with $W$  =1.8 $\mu$m, and period $L$ =4  $\mu$m (filling factor 0.45 or $L=2.2W$), whose raw data are reported in Fig. S1 of the SI. Incidentally, the agreement with the dispersion curve of the green diamond with L = 2.2W suggests that small changes in the filling factor do not affect the plasmon absorption whose characteristic frequency is mainly driven by W \cite{Martin}.
As shown in the inset, $\nu_{P} \propto W^{-1/2}$, as expected for a 2D plasmonic excitation. In the main panel of the Figure, the same points are plotted vs. $k$, based on the relation $k \simeq \pi/W$. Here, the approximation accounts both for a possible depletion in the electron density at the ribbon edges as observed in graphene \cite{Yan} and for scattering from the edges \cite{Yan}. As both phenomena affect only  a range  of tens of nanometers, smaller than our $W$ by two orders of magnitude, the corresponding correction can be neglected. The effective value of  $k$ may be also influenced by  the  excitation of edge modes \cite{Nikitin1,Nikitin2} as those predicted in graphene, but those results cannot be easily transferred to the present material where, for example, the electron mobility  is lower by three orders of magnitude than that considered in those calculations. In any case, by assuming $k =\pi/W$ we find that our points  are in good agreement with the dispersion law  $\nu_{P} \propto \sqrt{k}$ predicted by the Equations discussed below and reported by the black dashed line in Fig. \ref{dispersion}. 

Once established that the observed plasmon is that of the 2D electron gas at the TI surface, one wonders whether it should be ascribed mainly to the Dirac fermions or to the massive electrons. As we discuss in the following, both qualitative and quantitative arguments support the former assignment. 
The qualitative argument starts from the observation that the plasmon absorption linewidths in Fig. \ref{ribbons}-e are very similar at 6 K and 300 K (see also the Fano fit parameters in Tables S2 and S3 of the SI). This shows that the plasmon absorption is remarkably robust against a variation in temperature by a factor of 50. This would be hard to explain if a major contribution to the absorption were due to collective excitations of conventional electrons. Due to the momentum-spin lockage, the Dirac quasiparticles in TI's are instead  virtually unaffected by the scattering mechanisms  - except for those with magnetic impurities (here absent) - and therefore by their temperature dependence. Indeed, a weak broadening of TI's states with temperature has been observed in Bi$_2$Se$_3$ by Angle-Resolved Photoemission Spectroscopy  \cite{Valla}, consistently with the present results, even if the electron-phonon interaction is reported to be strong \cite{Zhu}.  This suggests that the TIs Dirac electrons are protected not only from impurity scattering but also from scattering by phonons. One may therefore speculate that the weak temperature dependence of the plasmon linewidth is related to the  topological protection of the electrons. The infrared absorption of "massive" plasmons in 2D electron gases was reported only below rather low temperatures \cite{Allen} because the electron-phonon scattering is their main decay channel \cite{Baumberg, Cho}. Those temperatures, moreover, decrease as the leading longitudinal optical phonon frequency $\hbar\omega_{LO}$ decreases: they are close to that of liquid nitrogen in GaN heterostructures with $\hbar\omega_{LO}$ = 92 meV \cite{Stanton}, to that of liquid helium in GaAs heterostructures with $\hbar\omega_{LO}$ = 36 meV \cite{Batke}. As in Bi$_2$Se$_3$  $\hbar\omega_{LO}$ $\sim$ 19 meV \cite{Paglione}, therein we should expect the linewidth of 2D massive plasmons being so large that their absorption is confused in the background, even at the lowest  $T$ (6 K).  A robustness vs. temperature of the plasmonic excitations was found also in graphene \cite{Martin}, where Dirac-plasmon absorption peaks were observed at 300 K. Therein, however, one has a much higher $\hbar\omega_{LO}$ = 137 meV.

The quantitative argument for the assignment of the plasmon peaks in Fig. 2 to Dirac quasiparticles proceeds as follows. In TI thin films, the  2D free-electron layers at the TI-substrate and at the TI-vacuum interfaces interact through an effective Coulomb potential \cite{Polini, DasSarma}. This mechanism, if one neglects interlayer tunneling, leads to the appearance of two longitudinal collective excitations, \textit{i. e.} an optical plasmon (with $\nu_{P} \propto \sqrt{k}$), and an acoustic plasmon (with $\nu_{P} \propto k$) \cite{Polini, DasSarma}.
The acoustic mode has been estimated \cite{Polini} to be degenerate with the  continuum, \textit{i. e.}, to be strongly Landau-damped \cite{Landau} and, then, unobservable.  Therefore, the THz spectrum of TI's should be characterized by a single optical mode to which, in principle, would participate both  massive and Dirac fermions. In the long wavelength limit  $k\rightarrow0$, the dispersion laws of Dirac and massive plasmons can be written, respectively \cite{Polini, DasSarma},

\begin{equation}
 \nu_{D}(k)= \frac{1}{2\pi}\sqrt{k}  \left(\frac{e^2}{4\pi\epsilon_0\epsilon\hbar}  v_F \sqrt{2\pi n_Dg_s g_v}\right)^{1/2}
 \label{Dirac}
\end{equation}

\noindent and

\begin{equation}
 \nu_M (k)= \frac{1}{2\pi}\sqrt{k}  \left(\frac{e^2}{2\epsilon_0\epsilon}n_M/m^*\right)^{1/2}
 \label{massive}
\end{equation}

\noindent
where the spin ($g_s$) and the valley  ($g_v$) degeneracies are both equal to 1 \cite{Polini}. Moreover, in Bi$_2$Se$_3$ thin films grown in the same conditions as the present ones, $n_{M} = 7.5\pm3.5\times10^{12}$ cm$^{-2}$, $n_{D}=3\pm1\times10^{13}$ cm$^{-2}$, $v_{F}= 6\pm1\times10^{7}$ cm/s  \cite{Cao}, and the effective mass of the parabolic bulk band is $m^*$ = 0.15$\pm$0.01 $m_e$ \cite{Bansal, Bansal1}. 
Neither Eq. \ref{Dirac} nor Eq. \ref{massive} contain the film thickness or  the bulk dielectric function. They instead depend on the average  $\epsilon$=($\epsilon_{1}$+$\epsilon_{2}$)/2  between the dielectric function of a vacuum ($\epsilon_{1}=1$) and that of the substrate ($\epsilon_{2}\sim10$). The theoretical Dirac and massive plasmon dispersions were calculated by Eqs. \ref{Dirac} and \ref{massive} at the selected wavevectors $k=\pi/W$ by using for $n_{D}$, $v_F$ (Dirac), $n_{M}$ and $m^*$ (massive) their central experimental values reported above. The results are  compared with the experimental data reported by dots in Fig. \ref{dispersion}. Therein,  the dashed black curve is the  dispersion predicted for a Dirac plasmon in Bi$_2$Se$_3$,  the dotted blue line that of massive particles. The much better  agreement with data of the Dirac plasmon dispersion with respect to the "massive" one, obtained without using  free parameters, strongly supports the above qualitative argument, namely  that the plasmonic excitations observed in this experiment must be ascribed to Dirac topological carriers.

In conclusion, we have reported the first observation of Dirac plasmon resonances in thin films of Bi$_2$Se$_3$. We have shown, based on both qualitative arguments and the comparison with theoretical calculations, that those plasmons should be assigned to Dirac quasi-particles at the metallic surface of the topological insulator. The strong Fano-like interference between the Dirac plasmons and the  phonons of Bi$_2$Se$_3$, here observed, opens promising perspectives. Indeed, similar interference effects observed in conventional plasmonic systems between bright (dipole) modes and dark (quadrupole) modes \cite{giannini} were used to increase the quality factor of resonances. Similar opportunities can be exploited in TI's, for example, to implement novel sensors in the terahertz range. Finally, the plasmon tunability, here engineered through microribbon arrays of different width $W$ and period $2W$, can be generalized to produce more complex plasmonic designs in these intriguing topological materials. 

\balancecolsandclearpage


\balancecolsandclearpage


\balancecolsandclearpage


\balancecolsandclearpage
\begin{widetext}

FIG 1: {\bf Extinction coefficients of the micro ribbon arrays of Bi$_2$Se$_3$ topological insulators in the terahertz range.} {\bf a.} Scanning Electron Microscope (SEM) image of the $W$=2.5 $\mu$m patterned film. {\bf b.} Extinction coefficient of the as-grown, unpatterned film, at 6 K (blue lines) and 300 K (red lines).  {\bf c.} Optical-microscope images of the five patterned films with different widths $W$ and periods $2W$; the red arrows indicate the direction of radiation polarization, either perpendicular or parallel to the ribbons. The film thickness is reported under the images. {\bf d.} Extinction coefficient at 6 K (blue lines) and 300 K (red lines) of the five patterned films, with the radiation field parallel to the ribbons. {\bf e.} Extinction coefficient of the five patterned films, with the radiation polarized perpendicularly to the ribbons, at 6 K (blue lines) and 300 K (red lines). All data are normalized by their respective peak values.\\
\\

FIG 2: {\bf Search for the bare plasmon frequencies.} Normalized extinction coefficient of the five patterned films, with the radiation polarized perpendicularly to the ribbons, at 6 K (circles) and fits to Eq. \ref{giannini_1} (black lines). The bare plasmon and  phonon contributions, extracted through the fits, are  reported by the red and the green line, respectively. The inset in the bottom panel displays the plasmon linewidth $\Gamma_P$ vs. the ribbon width $W$ at 6 K. The dotted line is the Drude contribution extracted from data with polarization parallel to the ribbons, the dashed line is a guide to the eye. Both in the top and bottom panel the bump at 2.6 THz is instrumental and due to a bad compensation of the Mylar beamsplitter absorption.\\
\\

FIG 3: {\bf Experimental and theoretical dispersion of plasmons in Bi$_2$Se$_3$.}  Inset:  linear dependence of $\nu_{P}$ on $W^{-1/2}$, where $W= \pi/k$ is the ribbon width. Main panel: experimental values at 6 K (blue full circles) compared with the plasmon dispersion for Dirac (dashed black line)  and massive electrons (dotted blue line) calculated with no fitting parameters by Eq. \ref{Dirac} and Eq. \ref{massive}, respectively. The green diamond refers to a sample with $W$  =1.8 $\mu$m and period $L=2.2W$, see text.\\
\\

\end{widetext}

\balancecolsandclearpage

\section{Methods}
The high quality Bi$_2$Se$_3$	 thin-films were prepared by MBE using the standard two-step growth method developed at Rutgers University \cite{Bansal, Bansal1}. The $10\times10$ mm$^{2}$ Al$_2$O$_3$ substrates were first cleaned by heating to 750 $^o$C in an oxygen environment to remove organic surface contamination. The substrates were then cooled to 110 $^o$C, where an initial 3 QL of Bi$_2$Se$_3$ was deposited. This was followed by heating to 220 $^o$C, where the remainder of the film was deposited to attain the target thickness. The Se and Bi flux ratio was kept to be approximately $Se/Bi$ $10/1$, to minimize Se vacancies. Once the films were cooled, they were removed from the vacuum chamber, and vacuum-sealed in plastic bags within two minutes, and shipped to the University of Rome.

Bi$_2$Se$_3$ ribbons were fabricated by electron-beam lithography (EBL) and subsequent Reactive Ion Etching (RIE). The Bi$_2$Se$_3$ film was spin-coated with a double layer of electron-sensitive resist polymer PMMA (Poly-(methyl methacrylate)) up to a total thickness of 1.4 microns. The ribbon pattern with different $W$ was then written in the resist by EBL. In order to obtain a lithographic pattern with re-alignment precision below 10 nm over a sample area suitable for Terahertz spectroscopy of 10x10 mm$^2$, we used an electron beam writer equipped with a XY interferometric stage (Vistec EBPG 5000). The patterned resist served as mask for the removal of Bi$_2$Se$_3$ by RIE at low microwave power of 45 W to prevent heating of the resist mask. 
Sulfur hexafluoride (SF$_6$) was used as the active reagent. The Bi$_2$Se$_3$ film was etched at a rate of 20 nm/min, which was verified by Atomic Force Microscopy (AFM) after soaking the sample in acetone to remove PMMA. The in-plane edge quality after the RIE process, as inspected by AFM, closely follows that of the resist polymer mask, i.e. edge roughness smaller than 20 nm. The vertical profile of the edge forms an angle of about 45 degrees with the substrate plane, because our RIE process has no preferred etching direction.

The absorption spectra in the Terahertz range were obtained by using a Bruker IFS-66V Michelson interferometer and a liquid-He cooled bolometer.  The Bi$_2$Se$_3$ film and a co-planar Al$_2$O$_3$ bare substrate were mounted on the cold finger of a He-flow cryostat and kept at a pressure of about 10$^{-6}$ mbar. The radiation was polarized either along, or perpendicular to, the ribbons by a THz polarizer having a degree of polarization $>$ 99.5 \%. The extinction coefficient  reported in Fig. 2 was obtained from the film transmittance $T$, defined as the ratio between the intensity transmitted by the thin film and that transmitted by the bare substrate. 

The Fano fits were obtained by replacing in Eq. \ref{giannini_1} \cite{giannini}

\begin{equation}
\nu'  =  \frac{\nu-\nu_{ph}}{\Gamma_{ph}(\nu)/2} - \frac{\nu - \nu_{P}}{\Gamma_{P}/2}
\label{giannini_2} 
\end{equation}

\noindent the plasmon-coupled phonon linewidth

\begin{equation}\label{calculated_3}
	\Gamma_{ph}(\nu) =  \frac{2 \pi v^2}{1+\left(\frac{\nu-\nu_{P}}{\Gamma_{P}/2}\right)^2}
\end{equation}

\noindent and the Fano factor

\begin{equation}
q(\nu) =  \frac{vw/g}{\Gamma_{ph}(\nu)/2} + \frac{\nu - \nu_{P}}{\Gamma_{P}/2} \\
\label{giannini_3}
\end{equation}

\noindent Therein, $w$ and $g$ are the coupling factors of the radiation with the phonon and the plasmon, respectively, $v$ measures the phonon-plasmon Fano interaction, and $\Gamma_{P}$ is the width of the plasmon line, assumed to be Lorentzian.

\section{Acknowledgements}
We thank M. Polini for fruithful discussions about Dirac and massive plasmonic dispersions.

\section{Authors Contributions} 
M. B., N. B., N. K. and S. O. fabricated and characterized Bi$_{2}$Se$_{3}$ films. M.O. and A. D.G. and V.G. performed EBL lithography and etching. P.D.P, F. G., O.L. and M. O. carried out the terahertz experiments and data analysis. P.C., M. O., and S. L. were responsible for the planning and the management of the project with inputs from all the co-authors, especially from P.D.P., F. G. S. O. and O.L. All authors extensively discussed the results and the manuscript that was written by P.C., M.O. and S.L. 
 
\section{Additional Information}
The authors declare no competing financial interests. Correspondence and requests for materials should be addressed to S.L. (stefano.lupi@roma1.infn.it) 

\balancecolsandclearpage


\section{References}
\begin{thebibliography}{99}

\bibitem{Maier}
\bibinfo{author}{Maier S. A.}
\newblock \bibinfo{title}{\textit{Plasmonics: fundamentals and applications}}.
\newblock \emph{\bibinfo{journal}{Springer-Verlag, New York,}}
(\bibinfo{year}{2007}).
  
 \bibitem{Martin}
\bibinfo{author}{Ju, L.} \emph{et al.}
\newblock \bibinfo{title}{Graphene plasmonics for tunable Terahertz
  metamaterials}.
\newblock \emph{\bibinfo{journal}{Nat Nano}} \textbf{\bibinfo{volume}{6}},
  \bibinfo{pages}{630} (\bibinfo{year}{2011}).
  
 \bibitem{Grigorenko}
\bibinfo{author}{Grigorenko, A. N.} \bibinfo{author}{Polini, M.} \&
  \bibinfo{author}{Novoselov, K. S.}
\newblock \bibinfo{title}{Graphene plasmonics}.
\newblock \emph{\bibinfo{journal}{Nat Phot}} \textbf{\bibinfo{volume}{6}},
  \bibinfo{pages}{749} (\bibinfo{year}{2012}).

    
\bibitem{Hasan1}
\bibinfo{author}{Hasan, M. Z.} \& \bibinfo{author}{Kane, C.~L.}
\newblock \bibinfo{title}{\textit{Colloquium} : Topological insulators}.
\newblock \emph{\bibinfo{journal}{Rev. Mod. Phys.}}
  \textbf{\bibinfo{volume}{82}}, \bibinfo{pages}{3045--3067}
  (\bibinfo{year}{2010}).

\bibitem{Kane1}
\bibinfo{author}{Kane, C. L.} \& \bibinfo{author}{Mele, E.~J.}
\newblock \bibinfo{title}{Quantum spin hall effect in graphene}.
\newblock \emph{\bibinfo{journal}{Phys. Rev. Lett.}}
  \textbf{\bibinfo{volume}{95}}, \bibinfo{pages}{226801}
  (\bibinfo{year}{2005}).

\bibitem{Moore1}
\bibinfo{author}{Moore, J. E.}
\newblock \bibinfo{title}{The birth of topological insulators}.
\newblock \emph{\bibinfo{journal}{Nature}} \textbf{\bibinfo{volume}{464}},
  \bibinfo{pages}{194} (\bibinfo{year}{2010}).

\bibitem{Qi}
\bibinfo{author}{Qi, X.-L.}, \bibinfo{author}{Hughes, T.~L.} \&
  \bibinfo{author}{Zhang, S.-C.}
\newblock \bibinfo{title}{Topological field theory of time-reversal invariant
  insulators}.
\newblock \emph{\bibinfo{journal}{Phys. Rev. B}} \textbf{\bibinfo{volume}{78}},
  \bibinfo{pages}{195424} (\bibinfo{year}{2008}).

\bibitem{Essin}
\bibinfo{author}{Essin, A. M.}, \bibinfo{author}{Moore, J.~E.} \&
  \bibinfo{author}{Vanderbilt, D.}
\newblock \bibinfo{title}{Magnetoelectric polarizability and axion
  electrodynamics in crystalline insulators}.
\newblock \emph{\bibinfo{journal}{Phys. Rev. Lett.}}
  \textbf{\bibinfo{volume}{102}}, \bibinfo{pages}{146805}
  (\bibinfo{year}{2009}).
  
  \bibitem{Fu}
\bibinfo{author}{Fu, L.} \& \bibinfo{author}{Kane, C. L.}
\newblock \bibinfo{title}{Superconducting proximity effect and majorana
  fermions at the surface of a topological insulator}.
\newblock \emph{\bibinfo{journal}{Phys. Rev. Lett.}}
  \textbf{\bibinfo{volume}{100}}, \bibinfo{pages}{096407}
  (\bibinfo{year}{2008}).

\bibitem{Nilsonn}
\bibinfo{author}{Akhmerov, A. R.}, \bibinfo{author}{Nilsson, J.} \&
  \bibinfo{author}{Beenakker, C. W. J.}
\newblock \bibinfo{title}{Electrically detected interferometry of majorana
  fermions in a topological insulator}.
\newblock \emph{\bibinfo{journal}{Phys. Rev. Lett.}}
  \textbf{\bibinfo{volume}{102}}, \bibinfo{pages}{216404}
  (\bibinfo{year}{2009}).

\bibitem{Fu1}
\bibinfo{author}{Fu, L.} \& \bibinfo{author}{Collins, G. P.}
\newblock \bibinfo{title}{Computing with quantum knots}.
\newblock \emph{\bibinfo{journal}{Sci. Am.}} \textbf{\bibinfo{volume}{294}},
  \bibinfo{pages}{57} (\bibinfo{year}{2006}).

\bibitem{Kitaev1}
\bibinfo{author}{Kitaev, A.} \& \bibinfo{author}{Preskill, J.}
\newblock \bibinfo{title}{Topological entanglement entropy}.
\newblock \emph{\bibinfo{journal}{Phys. Rev. Lett.}}
  \textbf{\bibinfo{volume}{96}}, \bibinfo{pages}{110404}
  (\bibinfo{year}{2006}).

\bibitem{zangh}
\bibinfo{author}{Zhang, X.}, \bibinfo{author}{Wang, J.} \&
  \bibinfo{author}{Zhang, S.-C.}
\newblock \bibinfo{title}{Topological insulators for high-performance Terahertz
  to infrared applications}.
\newblock \emph{\bibinfo{journal}{Phys. Rev. B}} \textbf{\bibinfo{volume}{82}},
  \bibinfo{pages}{245107} (\bibinfo{year}{2010}).

\bibitem{Chen}
\bibinfo{author}{Chen, Y.} \emph{et al.}
\newblock \bibinfo{title}{Experimental realization of a three-dimensional
  topological insulator, ${\mathrm{Bi}}_{2}{\mathrm{Te}}_{3}$}.
\newblock \emph{\bibinfo{journal}{Science}} \textbf{\bibinfo{volume}{325}},
  \bibinfo{pages}{178} (\bibinfo{year}{2009}).
  
  \bibitem{Allen}
\bibinfo{author}{Allen, S. J.}, \bibinfo{author}{Tsui, D. C.} \& \bibinfo{author}{Logan R. A.}
\newblock \bibinfo{title}{Observation of the two-Dimensional plasmon in Silicon Inversion Layers}.
\newblock \emph{\bibinfo{journal}{Phys. Rev. Lett.}}
  \textbf{\bibinfo{volume}{38}}, \bibinfo{pages}{980--983}
  (\bibinfo{year}{1977}).

 \bibitem{Nikitin1} 
\bibinfo{author}{Nikitin A. Yu.},\bibinfo{Guinea F.}{Garcia Vidal F.J.} \& \bibinfo{author}{Martin Moreno L.}
newblock \bibinfo{title}{Edge and waveguide terahertz surface plasmon modes in graphene microribbons}.
\newblock \emph{\bibinfo{journal}{Phys. Rev. B}}
  \textbf{\bibinfo{volume}{84}}, \bibinfo{pages}{1614071-1614074}
  (\bibinfo{year}{2011}).


 \bibitem{Nikitin2}
\bibinfo{author}{Nikitin A. Yu.},\bibinfo{Guinea F.}{Garcia Vidal F.J.} \& \bibinfo{author}{Martin Moreno L.}
newblock \bibinfo{title}{Surface plasmon enhanced absorption and suppressed transmission in periodic arrays of graphene ribbons}.
\newblock \emph{\bibinfo{journal}{Phys. Rev. B}}
  \textbf{\bibinfo{volume}{85}}, \bibinfo{pages}{081405(R)1-081405(R)4}
  (\bibinfo{year}{2012}).


 \bibitem{Yan} 
 \bibinfo{author}{Yan H.}\emph{et al.}
\newblock \bibinfo{title}{Damping pathways of mid-infrared plasmons in graphene nanostructures}.
\newblock \emph{\bibinfo{journal}{Nature Photon.}} \text{\bibinfo{volume}{doi:10.1038/nphoton.2013.57}},
 (\bibinfo{year}{2013}).


\bibitem{Polini}
\bibinfo{author}{Profumo, R.E.V.} \emph{et al.}
\newblock \bibinfo{title}{Double-layer graphene and topological insulator thin-film plasmons}.
\newblock \emph{\bibinfo{journal}{Phys. Rev. B}},
  \textbf{\bibinfo{volume}{85}}, \bibinfo{pages}{085443}
  (\bibinfo{year}{2012}).
  
  \bibitem{DasSarma}
\bibinfo{author}{Das~Sarma, S.} \& \bibinfo{author}{Hwang, E.~H.}
\newblock \bibinfo{title}{Collective modes of the massless Dirac plasma}.
\newblock \emph{\bibinfo{journal}{Phys. Rev. Lett.}}
  \textbf{\bibinfo{volume}{102}}, \bibinfo{pages}{206412}
  (\bibinfo{year}{2009}).
   
\bibitem{Bansal}
\bibinfo{author}{Bansal, N.}, \bibinfo{author}{Kim, Y. S.},
  \bibinfo{author}{Brahlek, M.}, \bibinfo{author}{Edrey, E.} \&
  \bibinfo{author}{Oh, S.}
\newblock \bibinfo{title}{Thickness-independent transport channels in
  topological insulator ${\mathrm{Bi}}_{2}{\mathrm{Se}}_{3}$ thin films}.
\newblock \emph{\bibinfo{journal}{Phys. Rev. Lett.}}
  \textbf{\bibinfo{volume}{109}}, \bibinfo{pages}{116804}
  (\bibinfo{year}{2012}).

\bibitem{Bansal1}
\bibinfo{author}{Bansal, N.}, \emph{et al.},
\newblock \bibinfo{title}{Epitaxial growth of topological insulator ${\mathrm{Bi}}_{2}{\mathrm{Se}}_{3}$ thin film on Si(111) with atomically sharp interface}.
\newblock \emph{\bibinfo{journal}{Thin Solid Film}}
  \textbf{\bibinfo{volume}{520,}}, \bibinfo{pages}{224}
  (\bibinfo{year}{2011}).

\bibitem{Dipietro}
\bibinfo{author}{Di~Pietro, P.} \emph{et al.}
\newblock \bibinfo{title}{Optical conductivity of bismuth-based topological
  insulators}.
\newblock \emph{\bibinfo{journal}{Phys. Rev. B}} \textbf{\bibinfo{volume}{86}},
  \bibinfo{pages}{4701} (\bibinfo{year}{2012}).

\bibitem{Armitage}
\bibinfo{author}{Valdes~Aguilar, R.}, \emph{et al.},
\newblock \bibinfo{title}{THz response and colossal Kerr rotation from the surface states of the topological insulator Bi$_2$Se$_3$}.
\newblock \emph{\bibinfo{journal}{Phys. Rev. Lett.}}
  \textbf{\bibinfo{volume}{108}}, \bibinfo{pages}{087403}
  (\bibinfo{year}{2012}).

 \bibitem{Basov}
\bibinfo{author}{Fei, Zhe} \emph{et al.}
\newblock \bibinfo{title}{Infrared Nanoscopy of Dirac Plasmons at the Graphene-SiO$_2$ interface}.
\newblock \emph{\bibinfo{journal}{Nano Lett.}} \textbf{\bibinfo{volume}{11}},
  \bibinfo{pages}{4701} (\bibinfo{year}{2011}).

\bibitem{giannini}
\bibinfo{author}{Giannini, V.}, \bibinfo{author}{Francescato, Y.},
  \bibinfo{author}{Amrania, H.}, \bibinfo{author}{Phillips, C. C.} \&
  \bibinfo{author}{Maier, S.~A.}
\newblock \bibinfo{title}{Fano resonances in nanoscale plasmonic systems: A
  parameter-free modeling approach}.
\newblock \emph{\bibinfo{journal}{Nano Lett.}} \textbf{\bibinfo{volume}{11}},
  \bibinfo{pages}{2835} (\bibinfo{year}{2011}).
  
\bibitem{Valla}
\bibinfo{author}{Pan, Z. -H}, \emph{et al.},
\newblock \bibinfo{title}{Measurement of an exceptionally weak electron-phonon coupling on the surface of the topological insulator Bi$_2$Se$_3$ using angle-resolved photoemission spectroscopy}.
\newblock \emph{\bibinfo{journal}{Phys. Rev. Lett.}}
 \textbf{\bibinfo{volume}{108}}, \bibinfo{pages}{187001}
 (\bibinfo{year}{2012}).
  
    
\bibitem{Zhu}
\bibinfo{author}{Zhu Xuetao},\emph{et al.},
\newblock \bibinfo{title}{Electron-Phonon Coupling on the Surface of the Topological Insulator Bi$_2$Se$_3$ Determined from
Surface-Phonon Dispersion Measurements}.
\newblock \emph{\bibinfo{journal}{Phys. Rev. Lett.}}
 \textbf{\bibinfo{volume}{108}}, \bibinfo{pages}{1855011}
 (\bibinfo{year}{2012}).

\bibitem{Baumberg}
\bibinfo{author}{Baumberg, J. J.} \& \bibinfo{author}{Williams D. A.}
\newblock \bibinfo{title}{Coherent phonon-plasmon modes in GaAs:Al$_x$Ga$_{1-x}$As heterostructures}.
\newblock \emph{\bibinfo{journal}{Phys. Rev. B}}
  \textbf{\bibinfo{volume}{53}}, \bibinfo{pages}{R16140--R16143}
  (\bibinfo{year}{1996}).

\bibitem{Cho}
\bibinfo{author}{Cho, G. C.} \bibinfo{author}{Dekorsy T.} \bibinfo{author}{Bakker H. J.} \bibinfo{author}{Hovel R.} \& \bibinfo{author}{Kurz H.}
\newblock \bibinfo{title}{Generation and relaxation of coherent majority plasmons}.
\newblock \emph{\bibinfo{journal}{Phys. Rev. Lett.}}
  \textbf{\bibinfo{volume}{77}}, \bibinfo{pages}{4062-4065}
  (\bibinfo{year}{1996}).
  
 \bibitem{Stanton}
\bibinfo{author}{Stanton, N. -M}, \emph{et al.},
\newblock \bibinfo{title}{Energy relaxation by hot electrons in n-GaN epilayers}.
\newblock \emph{\bibinfo{journal}{J. App. Phys.}}
 \textbf{\bibinfo{volume}{89}}, \bibinfo{pages}{973}
 (\bibinfo{year}{2001}). 

\bibitem{Batke}
\bibinfo{author}{Batke, E.} \bibinfo{author}{Heitmann D.} \& \bibinfo{author}{Wu C.}
\newblock \bibinfo{title}{Plasmon and magnetoplasmon excitation in two-dimensional electron space-charge layers on GaAs}.
\newblock \emph{\bibinfo{journal}{Phys. Rev. B.}}
  \textbf{\bibinfo{volume}{34}}, \bibinfo{pages}{6951--6960}
  (\bibinfo{year}{1986}).

\bibitem{Paglione}
\bibinfo{author}{A.B. Sushkov,} \emph{et al.,}
\newblock \bibinfo{title}{Far infrared cyclotron resonance and Faraday effect in low-doped Bi$_2$Se$_3$}.
\newblock \emph{\bibinfo{journal}{Phys. Rev. B}} \textbf{\bibinfo{volume}{82}},
 \bibinfo{pages}{125110} (\bibinfo{year}{2010}).
  
 \bibitem{Landau}
 \bibinfo{author}{Landau, L. } 
 \newblock \bibinfo{title}{On the vibration of the electronic plasma}.
\newblock \emph{\bibinfo{journal}{J. Phys.}} \textbf{\bibinfo{volume}{USSR 10}},
  \bibinfo{pages}{25} (\bibinfo{year}{1946}).
  
 \bibitem{Cao}
  \bibinfo{author}{Cao Yue}, \emph{et al.}
\newblock \bibinfo{title}{In-Plane Helical Orbital Texture Switch near the Dirac Point in the Topological Insulator ${\mathrm{Bi}}_{2}{\mathrm{Se}}_{3}$}.
\newblock \emph{\bibinfo{journal}{http://arxiv.org/abs/1209.1016}}
  
\end{thebibliography}
\end{document}